\documentclass[12pt]{iopart}

\usepackage{graphicx}
\usepackage{cite}
\usepackage{iopams}
\usepackage{bm}

\begin{document}

\title{Production of heavy neutron-rich nuclei in transfer reactions within the dinuclear system model}

\author{Long Zhu$^{1,2,}$\footnote{Email: zhulong@mail.bnu.edu.cn}, Zhao-Qing Feng$^{3,}$\footnote{Email: fengzhq@impcas.ac.cn}, and Feng-Shou Zhang$^{1,2,4,}$\footnote{Corresponding author: fszhang@bnu.edu.cn}}

\address{$^{1}$The Key Laboratory of Beam Technology and Material Modification of Ministry of Education, College of Nuclear Science and Technology, Beijing Normal University, Beijing 100875, China\\
$^{2}$Beijing Radiation Center, Beijing 100875, China\\
$^{3}$Institute of Modern Physics, Chinese Academy of Sciences, Lanzhou 730000, China\\
$^{4}$Center of Theoretical Nuclear Physics, National
Laboratory of Heavy Ion Accelerator of Lanzhou, Lanzhou 730000,
China\\}
\begin{abstract}
The dynamics of nucleon transfer processes in heavy-ion collisions is investigated within the dinuclear system model. The production cross sections of nuclei in the reactions $^{136}$Xe+$^{208}$Pb and $^{238}$U+$^{248}$Cm are calculated and the calculations are in good agreement with the experimental data. The transfer cross sections for the $^{58}$Ni+$^{208}$Pb reaction are calculated and compared with the experimental data. We predict the production cross sections of neutron-rich nuclei $^{165-168}$Eu, $^{169-173}$Tb, $^{173-178}$Ho, and $^{181-185}$Yb based on the reaction $^{176}$Yb+$^{238}$U. It can be seen that the production cross sections of the neutron-rich nuclei $^{165}$Eu, $^{169}$Tb, $^{173}$Ho, and $^{181}$Yb are 2.84 $\mu$b, 6.90 $\mu$b, 46.24 $\mu$b, and 53.61 $\mu$b, respectively, which could be synthesized in experiment.
\end{abstract}

\maketitle

\section{\label{int}Introduction}
The production of superheavy nuclei (SHN) have been studied extensively in the past three decades. Much progress has been made experimentally \cite{Hofmann01,Oganessian01,Kosuke01,Khuyagbaatar01} and theoretically \cite{Siwek01,Nan01,Shen01,Abe01,Mandaglio01,Zhu02,Wong01}. There are mainly two sorts of reaction mechanism to produce SHN, which are fusion reactions and multinulcleon transfer reactions. The synthesis of superheavy nuclei is motivated with respect to searching the ``island of stability". The heavy nuclei produced in fusion reactions are not so neutron-rich. Therefore, it is quite hard to reach the center of the ``island of stability" through the fusion process. The damped collisions between heavy nuclei could be one possible approach.

The light and medium mass nuclei far from the stability line (neutron- and proton-rich) can be produced in fragmentation process or fission of heavy nuclei. The heavy neutron-rich nuclei including those located at the ``island of stability" are not synthesized and studied yet, which could be synthesized by using multinucleon transfer reactions.

The fundamental mechanisms of the deep inelastic collisions have been investigated many years ago \cite{Cassing01,Norenberg01}. In recent years, the production of neutron-rich heavy and superheavy nuclei have been discussed by Zagrebaev and Greiner \cite{Zagrebaev01,Zagrebaev02}. The multinucleon transfer processes are also investigated by the time-dependent Hartree-Fock (TDHF) approach \cite{Sekizawa01,Golabek01} and improved quantum molecular dynamics (ImQMD) model \cite{Tian01}. The dinuclear system (DNS) model has been successfully used in investigating mechanisms of the synthesis of SHN in fusion reactions \cite{Adamian02,Adamian04,Adamian05,Adamian06,Zhu01} and the transfer processes of the damped collisions \cite{Adamian01,Feng03,Kalandarov01}.

In this work, the transfer cross sections are calculated within the framework of DNS model and compared with the available experimental data. The cross sections of transferring nucleons from light to heavy fragments and from heavy to light fragments are both investigated. We predict the production cross sections of the unknown neutron-rich nuclei through the transfer reaction $^{176}$Yb+$^{238}$U.

The article is organized as follows. In Sec. \ref{model}, we describe the theoretical model. The results and discussion are presented in Sec. \ref{result}. We summarize the main results in Sec. \ref{summary}.

\section{\label{model}Model description}
The diffusion process is treated along proton and neutron degrees of freedom. The distribution probability is obtained by solving a set of master equations in the potential energy surface (PES) of the DNS. The time evolution of the mass asymmetry is described by the following master equation \cite{Zhu01}:
\begin{eqnarray}
\fl \frac{dP(Z_{1},N_{1},E_{1},t)}{dt}
=\sum_{Z_{1}^{'}}W_{Z_{1},N_{1};Z_{1}^{'},N_{1}}(t)[d_{Z_{1},N_{1}}P(Z_{1}^{'},N_{1},E_{1}^{'},t)
-d_{Z_{1}^{'},N_{1}}P(Z_{1},N_{1},E_{1},t)]\nonumber\\
\fl +\sum_{N_{1}^{'}}W_{Z_{1},N_{1};Z_{1},N_{1}^{'}}(t)[d_{Z_{1},N_{1}}P(Z_{1},N_{1}^{'},E_{1}^{'},t)
-d_{Z_{1},N_{1}^{'}}P(Z_{1},N_{1},E_{1},t)]\nonumber\\
-\Lambda_{qf}(\Theta(t))P(Z_{1},N_{1},E_{1},t).
\end{eqnarray}
Here $P(Z_{1},N_{1},E_{1},t)$ is the probability distribution function of the fragment with proton number $Z_{1}$ and neutron number $N_{1}$ with corresponding excitation energy $E_{1}$ at time $t$.  $W_{Z_{1},N_{1};Z_{1}^{'},N_{1}}=W_{Z_{1},N_{1};Z_{1},N_{1}^{'}}$ is the mean transition probability from the channel ($Z_{1}$, $N_{1}$, $E_{1}$) to ($Z_{1}^{'}$, $N_{1}$, $E_{1}^{'}$) [or ($Z_{1}$, $N_{1}$, $E_{1}$) to ($Z_{1}$, $N_{1}^{'}$, $E_{1}^{'}$)]. $d_{Z_{1},N_{1}}$ denotes the microscopic dimension corresponding to the macroscopic state ($Z_{1}$, $N_{1}$, $E_{1}$). The details of $W$ and $d_{Z_{1},N_{1}}$ are given in Ref. \cite{Feng05}. The sum is taken over all possible proton and neutron numbers that fragment $Z_{1}^{'}$ and $N_{1}^{'}$ may take, but only one nucleon transfer is considered in the model. $\Lambda_{qf}$ describes the quasifission rate.

The evolution of the DNS along the relative distance $R$ leads to QF of the DNS. The QF rate $\Lambda_{qf}(\Theta(t))$ can be treated with the one-dimensional Kramers rate \cite{Adamian01},
\begin{eqnarray}
\Lambda_{qf}(\Theta(t))=&\frac{\omega}{2\pi\omega^{B_{qf}}}[\sqrt{(\frac{\Gamma}{2\hbar})^{2}+(\omega^{B_{qf}})^{2}}-\frac{\Gamma}{2\hbar}]
\times exp[-\frac{B_{qf}(Z_{1}, N_{1})}{\Theta(t)}].
\end{eqnarray}
The QF rate exponentially depends on the QF barrier $B_{qf}(Z_{1}, N_{1})$. The local temperature $\Theta(t)$ is calculated by using Fermi-gas expression $\Theta=\sqrt{E_{1}/a}$ with the local excitation energy $E_{1}$ and the level-density parameter $a=A/12$ $\textrm{MeV}^{-1}$. The frequency $\omega^{B_{qf}}$ of the inverted harmonic oscillator approximates the potential $V$ in $R$ at the top of the quasifission barrier, and $\omega$ is the frequency of the harmonic oscillator approximating the potential in $R$ around the bottom of the pocket. The $\Gamma$ determines the friction coefficients. Here, $\Gamma=2.8$ MeV, $\hbar\omega^{B_{qf}}=2.0$ MeV, and $\hbar\omega=3.0$ MeV.

The local excitation energy is defined as
\begin{eqnarray}
E_{1}=E_{diss}-&(U(Z_{1}, N_{1}, Z_{2}, N_{2}, J)-U(Z_{P}, N_{P}, Z_{T}, N_{T}, J))\nonumber\\
&-\frac{(J-M)^{2}}{2\zeta_{rel}}-\frac{M^{2}}{2\zeta_{int}}.
\end{eqnarray}
Here, $U(Z_{1}, N_{1}, Z_{2}, N_{2})$ and $U(Z_{P}, N_{P}, Z_{T}, N_{T})$ are the driving potentials of fragment $A_{1}$ and the entrance point of the DNS. $E_{diss}$ is the excitation energy of the composite system, which is converted from the relative kinetic energy loss. $M$ denotes the intrinsic angular momentum derived from the dissipation of the relative angular momentum, and $\zeta_{int}$ is the corresponding moment of inertia. $J$ denotes the initial angular momentum. $\zeta_{rel}$ is the relative moment of inertia of the DNS, which is given by $\zeta_{rel}=\mu R^{2}_{m}$. $R_{m}$ is the distance of two colliding nuclei located at bottom of potential pocket. For heavy systems with no potential pocket the transfer processes take place at the touching configuration.

During the diffusion process, the relative kinetic energy will dissipate into the DNS system. The local excitation energy is determined by the excitation energy of DNS and the PES \cite{Zhu01}. The PES of DNS is given as
\begin{eqnarray}
\fl U(Z_{1}, N_{1}, Z_{2}, N_{2}, R, \beta_{1}, \beta_{2}, \theta_{1}, \theta_{2}, J)=
U_{LD}(Z_{1}, N_{1})+U_{LD}(Z_{2}, N_{2})-\nonumber\\U_{LD}(Z, N)-
V^{CN}_{rot}(J)+V_{CN}(Z_{1}, N_{1}, Z_{2}, N_{2}, R, \beta_{1}, \beta_{2}, \theta_{1}, \theta_{2}, J).
\end{eqnarray}
$U_{LD}(Z_{1}, N_{1})$, $U_{LD}(Z_{2}, N_{2})$ and $U_{LD}(Z, N)$ are the binding energies of the fragments $A_{i}$ and compound nucleus $A$, respectively. The value of $U(Z_{1}, N_{1}, Z_{2}, N_{2}, R, \beta_{1}, \beta_{2}, \theta_{1}, \theta_{2}, J)$ is normalized to the energy of the rotating compound nucleus by $U_{LD}(Z, N)+V^{CN}_{rot}(J)$. $V_{CN}$ is the interaction potential of two fragments, which depends on the deformation parameter and the orientation of the deformed fragments ($\theta_{1}$ and $\theta_{2}$). The details of $V_{CN}$ are given in Ref. \cite{Adamian01}. The distance R between the center of two fragments is chosen to be the value at $R_{m}$ or the touching configuration, in which the DNS is assumed to be formed.

The sharing of the excitation energy between primary fragments is assumed to be proportional to their masses.  The code GEMINI is used to treat the sequential statistical evaporation of excited fragments.
Subsequent de-excitation cascades of the excited fragments via emission of light particles (neutron, proton, and $\alpha$) and gamma-rays competing with fission process are taken into account, which lead to the final mass distribution of the reaction products.
\begin{figure}
\begin{center}
\includegraphics[width=12cm,angle=0]{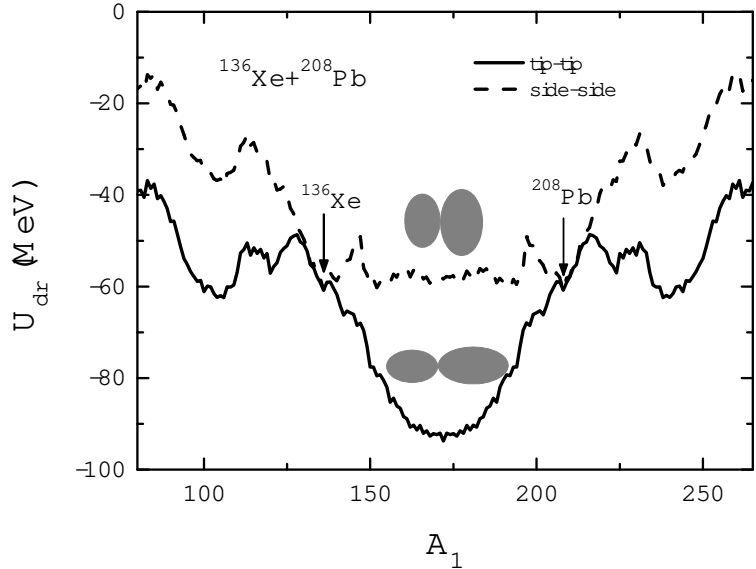}
\caption{\label{dri} Driving potentials of the tip-tip (solid line) and side-side (dashed line) collisions in the reaction $^{136}$Xe+$^{208}$Pb. }
\end{center}
\end{figure}

The production cross section of a primary fragment with charge $Z_{1}$ and mass number $A_{1}$ can be calculated as follows:
\begin{eqnarray}
\fl \sigma^{pr}_{Z_{1},A_{1}}(E_{c.m.})=\sum^{J_{max}}_{J=0}\sigma^{pr}_{Z_{1},A_{1}}(E_{c.m.},J)
=\sum^{J_{max}}_{J=0}\sigma_{cap}(E_{c.m.},J)P(Z_{1},A_{1},J,E_{1},\tau_{int}),
\end{eqnarray}
where $\sigma_{cap}(E_{c.m.},J)$ is the partial capture cross section, which defines the probability of overcoming the Coulomb barrier. $E_{c.m.}$ is the incident energy in center of mass system. The details of calculation of $\sigma_{cap}(E_{c.m.},J)$ is given in Ref. \cite{Zhu01}. $\tau_{int}$ is the interaction time of the diffusion process at the bottom of the potential pocket, which is dependent on the incident energy $E_{c.m.}$ and the incident angular momentum $J$. $\tau_{int}$ cannot be measured directly. In this work, $\tau_{int}$ is determined by using the deflection function method \cite{Li01}.

\section{\label{result}Results and discussion}

\begin{figure}
\begin{center}
\includegraphics[width=12cm,angle=0]{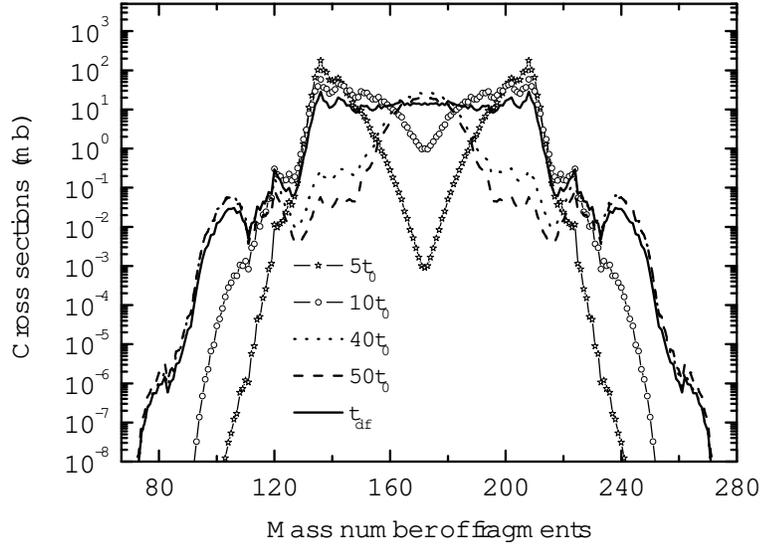}
\caption{\label{dis} The time evolution of the primary mass distributions of the reaction $^{136}$Xe+$^{208}$Pb at $E_{c.m.}=514$ MeV with $t_{0}=10^{-22}$ s. $t_{df}$ is the interaction time calculated by using deflection function method \cite{Li01}. }
\end{center}
\end{figure}

Figure \ref{dri} shows the driving potentials as a function of the mass number of fragments for two cases of tip-tip and side-side orientations in the reaction $^{136}$Xe+$^{208}$Pb. The arrows show the positions of the masses of projectile and target in entrance channel. A valley can be seen around the nucleus $^{208}$Pb for both orientations, which is because $^{208}$Pb is a double magic nucleus. It is also can be seen that the valley is deeper for side-side orientation. In this work, we investigate the dissipation process under the potential energy surface with tip-tip orientation.
\begin{figure}
\begin{center}
\includegraphics[width=12cm,angle=0]{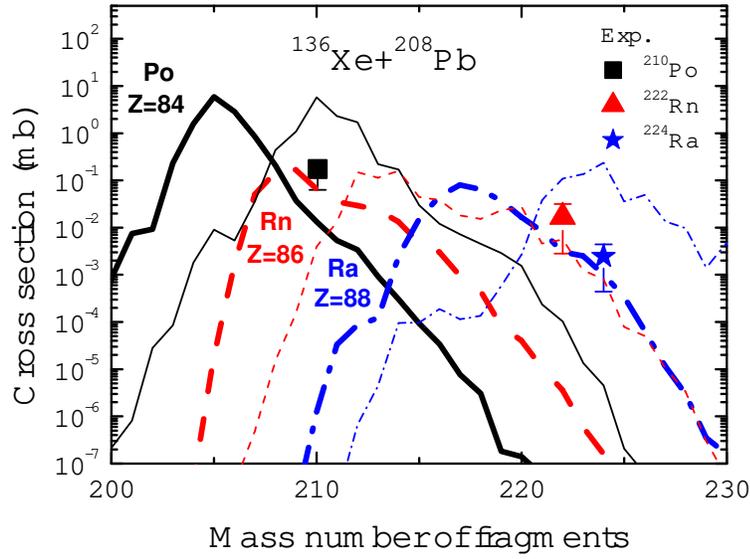}
\caption{\label{cross} The comparison of calculated production cross sections for primary (thin lines) and survived (thick lines) isotopes of Po (solid lines), Rn (dashed lines), and Ra (dash-dotted lines) with the experimental data \cite{Kozulin01} in the reaction $^{136}$Xe+$^{208}$Pb at $E_{c.m.}=514$ MeV. The solid points correspond to the experimental cross sections for $^{210}$Po (square), $^{222}$Rn (triangle), and $^{224}$Ra (star). }
\end{center}
\end{figure}

The time evolution of the primary mass distributions, for the reaction $^{136}$Xe+$^{208}$Pb at $E_{c.m.}=514$ MeV, is shown in Fig. \ref{dis}. It can be seen that the distribution is around the entrance channel with $\tau_{int}=5\times10^{-22}$ s. With the increasing interaction time the distributions become wider. Equilibrium is reached at about $\tau_{int}=40\times10^{-22}$ s. The distribution with $\tau_{int}=t_{df}$ is also shown. $t_{df}$ is calculated by using deflection function method. It is indicated that the system also can reach equilibrium with $\tau_{int}=t_{df}$, which can be seen from mass asymmetry region. The mass distribution around entrance channel for $\tau_{int}=t_{df}$ is higher than that for $\tau_{int}=40\times10^{-22}$ s. This is because $t_{df}$ depends on the angular momentum $J$. The interaction time decreases with the increasing angular momentum.
\begin{figure}
\begin{center}
\includegraphics[width=12cm,angle=0]{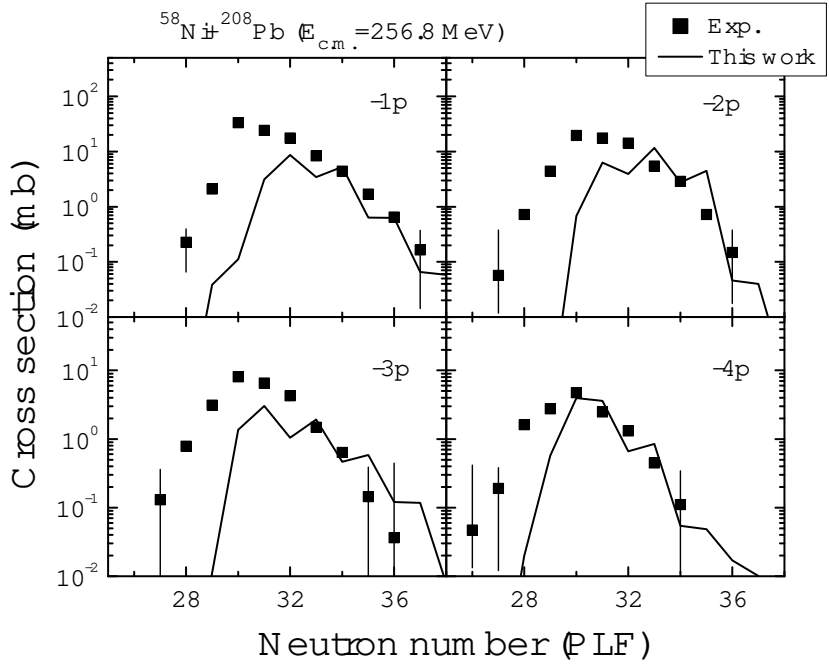}
\caption{\label{proton} Transfer cross sections for the $^{58}$Ni+$^{208}$Pb reaction at $E_{c.m.}=$256.8 MeV. The number of transferred protons from $^{58}$Ni to $^{208}$Pb (negative number) is indicated. The measured cross sections have been reported in Ref. \cite{Sekizawa01}.}
\end{center}
\end{figure}

In Fig. \ref{cross}, the calculated production cross sections of isotopes of Po, Rn, and Ra in transfer reaction $^{136}$Xe+$^{208}$Pb are shown. The bombarding energy $E_{c.m.}=$514 MeV. The calculated results are the case of tip-tip collisions, which have the height of the interaction potential at the touching configuration with the value 421 MeV. The larger qusifission rate of such systems with no potential pocket results in the DNS quickly decaying into two fragments. Therefore, the neutron transfer process is governed by the driving potential with short interaction time. Within the error bars, the calculated results are in good agreement with the experimental data \cite{Kozulin01} for production of $^{210}$Po and $^{224}$Ra, while the cross section of $^{222}$Rn are underestimated.
\begin{figure}
\begin{center}
\includegraphics[width=12cm,angle=0]{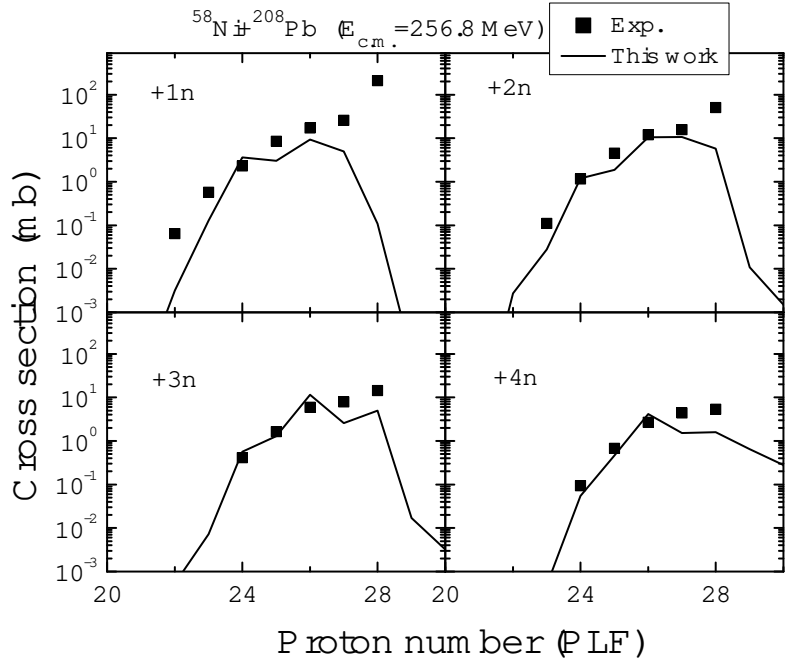}
\caption{\label{neutron} Transfer cross sections for the $^{58}$Ni+$^{208}$Pb reaction at $E_{c.m.}=$256.8 MeV. The number of transferred neutrons from $^{208}$Pb to $^{58}$Ni (positive number) is indicated. The measured cross sections have been reported in Ref. \cite{Sekizawa01}.}
\end{center}
\end{figure}

Figure \ref{proton} shows transfer cross sections of the reaction $^{58}$Ni+$^{208}$Pb at $E_{c.m.}=$256.8 MeV. Each panel of Fig. \ref{proton} shows cross sections according to the change of proton number of projectilelike fragment (PLF) from $^{58}$Ni, as functions of neutron number of PLF. The calculated results show a good agreement with the experimental data at neutron rich region, while the calculated results underestimate the experimental data at neutron lack region. As the transferred proton number decreases, the peak position of curve shifts towards larger neutron number, while the peaks of the experimental data locates at a neutron number of 30. The reason probably is that we only consider the driving potential with tip-tip orientation.
\begin{figure}
\begin{center}
\includegraphics[width=12cm,angle=0]{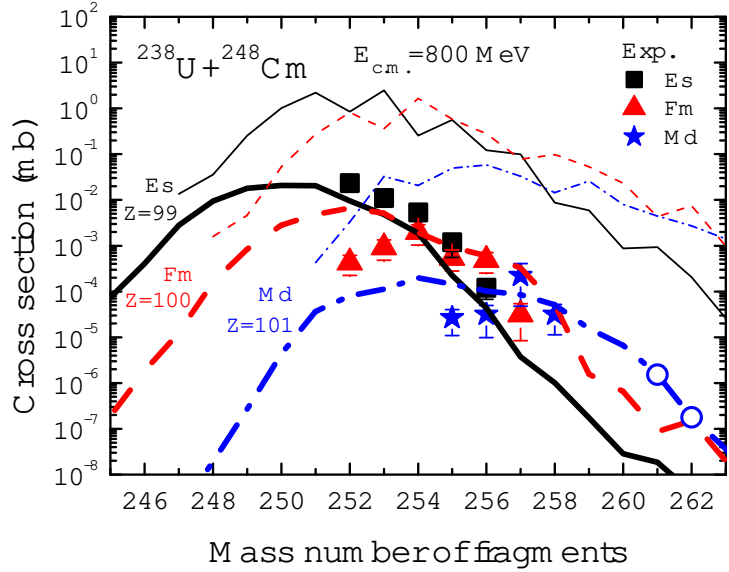}
\caption{\label{248Cm} Cross sections for the formation of isotopes of Einsteinium ($Z=99$) (solid lines), Fermium ($Z=100$) (dashed lines), and Mendelevium ($Z=101$) (dash-dotted lines) in the reaction $^{238}$U+$^{248}$Cm at $E_{c.m.}=800$ MeV. The thin and thick lines are distribution of primary and final fragments, respectively. The experimental data are taken from Ref. \cite{Schadel01}. The circles denote the unknown nuclei.}
\end{center}
\end{figure}

Figure \ref{neutron} shows transfer cross sections of the reaction $^{58}$Ni+$^{208}$Pb at $E_{c.m.}=$256.8 MeV according to the change of neutron number of PLF from $^{58}$Ni, as functions of proton number of PLF. From each panel of Fig. \ref{neutron} one can see that the calculated results are in good agreement with the experimental data except that the proton number is 28 in first panel. Therefore, the process of transferring some neutrons from $^{208}$Pb to $^{58}$Ni and transferring some protons from $^{58}$Ni to $^{208}$Pb can be better described.
\begin{figure}
\begin{center}
\includegraphics[width=12cm,angle=0]{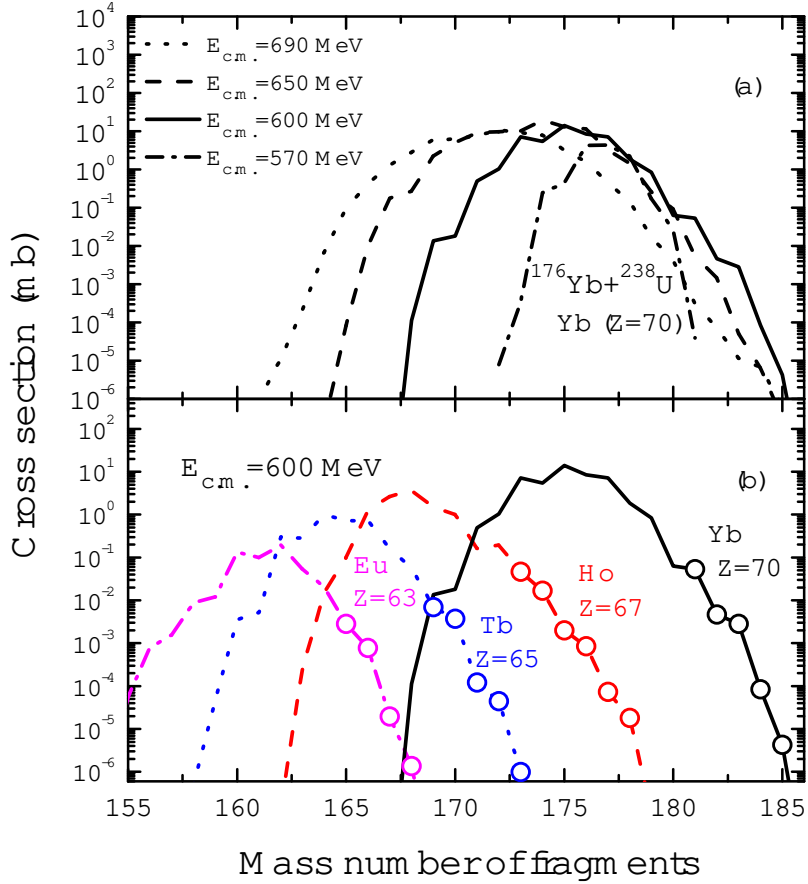}
\caption{\label{176Yb} (a) Production cross sections of isotopes of Yb in the transfer reaction $^{176}$Yb+$^{238}$U at $E_{c.m.}=570$ (dash-dotted line), 600 (solid line), 650 (dashed line), and 690 MeV (dotted line). (b) Cross sections for the formation of isotopes of the elements Ytterbium (solid line), Holmium (dashed line), Terbium (dotted line), and Europium (dash-dotted line) in the reaction $^{176}$Yb+$^{238}$U at $E_{c.m.}=600$ MeV. The circles denote the unknown neutron-rich nuclei.}
\end{center}
\end{figure}

In order to further test the model, the cross sections for the production of heavy actinides in damped collisions $^{238}$U+$^{248}$Cm at $E_{c.m.}=800$ MeV are shown in Fig. \ref{248Cm}. The interaction time of this system is very short due to the strong Coulomb repulsion. One can see a good agreement between theoretical and experimental cross sections \cite{Schadel01} for production of isotopes of the elements Es, Fm, and Md. The cross sections decrease drastically with increasing atomic numbers of the fragments. The nuclei $^{261}$Md and $^{262}$Md could be synthesized in this reaction with cross sections of about 1.52 nb and 0.17 nb, respectively. Reaction fragments formed in transfer reactions of heavy ions are strongly excited. It can be seen that the survival probabilities of most of primary fragments are quite low due to a dominant fission channel and the cross sections of primary fragments are a few hundreds times larger than those of the final fragments. The incident energy should also satisfy the condition that two colliding heavy nuclei can come into contact and have enough interaction time for nucleon transfer. Therefore, the optimal incident energy should be found for the largest yield of heavy neutron-rich nuclei.

In Fig. \ref{176Yb} the calculated yields of heavy neutron-rich nuclei in the transfer reaction $^{176}$Yb+$^{238}$U are shown. Figure \ref{176Yb} (a) shows the distributions of isotopes of the element Yb formed in the reaction $^{176}$Yb+$^{238}$U at different incident energies. The height of the interaction potential at the touching configuration is about 565 MeV. The height of curve increases with the increasing incident energy from $E_{c.m.}=570$ to 650 MeV, while it decreases from $E_{c.m.}=650$ to 690 MeV. This is because with the increasing incident energy, the excitation energy of these fragments increases and thus decreases their survival probabilities. It can be seen that cross sections are relatively larger in neutron-rich region when $E_{c.m.}=600$ MeV.

The calculated cross sections for production of Europium ($Z=63$), Terbium ($Z=65$), Holmium ($Z=67$), and Ytterbium ($Z=70$) isotopes in the damped collision $^{176}$Yb+$^{238}$U at $E_{c.m.}=600$ MeV are shown in Fig. \ref{176Yb} (b). The circles denote the unknown heavy neutron-rich nuclei. For the production of neutron-rich nuclei, the cross sections decrease dramatically with the increasing neutron number. We show in Fig. \ref{176Yb} (b) the neutron-rich nuclei with production cross section larger than 1 nb. The calculated production cross sections for the neutron-rich nuclei $^{165,166,167,168}$Eu are 2.84 $\mu$b, 0.78 $\mu$b, 19.64 nb, and 1.36 nb, $^{169,170,171,172,173}$Tb are 6.90 $\mu$b, 3.70 $\mu$b, 0.12 $\mu$b, 44.44 nb, and 1.00 nb, $^{173,174,175,176,177,178}$Ho are 46.24 $\mu$b, 16.83 $\mu$b, 2.00 $\mu$b, 0.85 $\mu$b, 73.75 nb, and 18.08 nb, and $^{181,182,183,184,185}$Yb are 53.61 $\mu$b, 4.67 $\mu$b, 2.85 $\mu$b, 84.70 nb, and 4.29 nb, respectively. One can see that the cross sections for synthesis of the heavy neutron-rich nuclei $^{165}$Eu, $^{169}$Tb, $^{173}$Ho, and $^{181}$Yb are 2.84 $\mu$b, 6.90 $\mu$b, 46.24 $\mu$b, and 53.61 $\mu$b, respectively, which are quite large for experimental detection.
\section{\label{summary}Conclusions}
The calculated production cross sections of heavy nuclei in transfer reaction $^{136}$Xe+$^{208}$Pb are in good agreement with the available experimental data. The transfer cross sections of the reaction $^{58}$Ni+$^{208}$Pb, according to the change of neutron (proton) number of PLF from $^{58}$Ni, as functions of proton (neutron) number of PLF are investigated. For the production of heavy neutron-rich isotopes of actinides, the reaction $^{238}$U+$^{248}$Cm is also studied. The calculated transfer cross sections show good agreement with the experimental data. It is reasonable to use the DNS model to study the dynamics of transfer reactions. By using the transfer reaction $^{176}$Yb+$^{238}$U, the neutron-rich nuclei $^{165}$Eu, $^{169}$Tb, $^{173}$Ho, and $^{181}$Yb could be synthesized with the cross sections of 2.84 $\mu$b, 6.90 $\mu$b, 46.24 $\mu$b, and 53.61 $\mu$b, respectively.

\section*{Acknowledgments}
This work was supported by the National Natural Science Foundation of China under Grand Nos. 11025524 and 11161130520, National Basic Research Program of China under Grant No. 2010CB832903, the European Commission's 7th Framework Programme (FP7-PEOPLE-2010-IRSES) under Grant Agreement Project No. 269131, and the National Natural Science Foundation of China Projects (Nos 11175218 and U1332207).

\end{document}